\documentclass[prd,twocolumn,aps,floats]{revtex4}
\usepackage{graphicx}
\usepackage{bm}
\usepackage{epsfig}
\usepackage{amsmath,amsfonts}
\usepackage{url}

\begin{document}

\title{Test particle motion in modified gravity theories}
\author{Mahmood Roshan}
\affiliation{School of Physics, Institute for Research in Fundamental Sciences (IPM), P. O. Box 19395-5531, Tehran, Iran}
\begin{abstract}
We derive the equations of motion of an electrically neutral test particle for modified gravity theories for which the covariant divergence of the ordinary matter energy-momentum tensor does not vanish (i.e. $\nabla_{\mu}T^{\mu\nu}\neq0$ ). In fact, we generalize the Mathisson-Papapetrou equations by deriving a general form for the equations of motion of a test particle. Furthermore, using the generalized Mathisson-Papapetrou equations, we investigate the equations of motion of a pole-dipole (spinning) particle in the context of Modified Gravity (MOG).
\end{abstract}
\maketitle \section{Introduction}
Deriving the equations of motion of a test particle from field equations of general relativity has a long history, see, for example, the account in \cite{review}. It is now well established in the literature that the equations of motion need not to be postulated separately,
but can be derived from the field equations of general relativity. More precisely, the conservation of the ordinary matter energy-momentum tensor, i.e. $\nabla_{\mu}T^{\mu\nu}=0$, is enough to find the geodesic equation and also the equations of motion of a spinning test particle (pole-dipole particle). In general relativity, a pole-dipole particle is described by the four-momentum and the tensor of spin, and the dynamics is governed by the Mathisson-Papapetrou equations \cite{mat,papa}. Combining these equations with a proper supplementary condition, one gets a self-consistent set of equations for describing the motion of a pole-dipole particle in a given space-time, for example see \cite{Suzuki:1996gm, appof,Semerak:1999qc,Mohseni:2000re,Obukhov:2010kn}. The case in which both gravitational and electromagnetic
fields are present and the particle is charged was studied by Dixon and Souriau \cite{dixon, souriau}. The equations of motion of a charged spinning test particle are known as Dixon-Souriau equations. For some solutions of the Dixon-Souriau equations in the specific backgrounds, see \cite{hojman, prasanna, ruffini}.

The multipole approximation method of \cite{mat,papa} has been used to investigate the motion of a test particle in the context of some modified gravity theories, for example see \cite{Yasskin:1980bu,mofijun,Puetzfeld:2008xu}. In this paper, using the multipole method, we derive the equations of motion of a pole-dipole test particle in modified gravity theories in which the covariant divergence of the ordinary matter energy-momentum tensor is given by $\nabla_{\mu}T^{\mu\nu}=A^{\nu}$, where the nonminimal coupling term $A^{\nu}$ is an arbitrary covariant vector, and in each modified gravity theory this vector can be determined exactly when the particular way by which the gravitational fields of the theory couple with the ordinary matter is specified. It is obvious that $A^{\nu}$ leads to a modification of the equation of motion of the test particle. In fact, we generalize the Mathisson-Papapetrou equations in such a way that they can be used in any torsionless non-metric theory of gravity. With metric theories of gravity we mean theories which postulate that (i) space-time is endowed with a metric $g_{\mu\nu}$ (ii) the world lines of single-pole particles are geodesics of that metric, and (iii) in the freely falling frames, the non-gravitational laws of physics are those of special relativity \cite{Willbook}. It should be noted that, in metric theories of gravity, the ordinary matter energy-momentum tensor is conserved, the non-metricity $Q_{\mu\nu\lambda}=-\nabla_{\mu}g_{\nu\lambda}$ is zero, and consequently the single-pole particles motion is described by the geodesic equation and the pole-dipole particles motion is governed by the Mathisson-Papapetrou equations. Furthermore, we make use of the generalized Mathisson-Papapetrou equations in order to investigate the test particle equations of motion in Modified Gravity (MOG) \cite{mofi}. In addition, we derive some conserved quantities for the test particle motion.
\section{Energy-momentum conservation }
Let us study the test particle motion in a modified gravity theory without torsion in which the usual conservation law of the ordinary matter energy-momentum tensor is replaced by
\begin{equation}
\nabla_{\mu}T^{\mu\lambda}=A^{\lambda}
\label{yek}
\end{equation}
We know that modified gravity theories which yield to such an identity for the energy-momentum tensor, would violate the Einstein equivalence principle. In fact, one can show that the covariant divergence of the energy-momentum tensor is zero if the gravitational fields (such as scalar fields or vector fields) of the theory do not interact directly with the matter. In the presence of the test particle, the components of the metric tensor are $\hat{g}_{\mu\lambda}=g_{\mu\lambda}+\delta g_{\mu\lambda}$, where $g_{\mu\lambda}$ is the background metric and $\delta g_{\mu\lambda}$ is the metric perturbation caused by the test particle, and it is supposed to be small. Also, let the energy-momentum tensor be $\hat{T}^{\mu\lambda}=T^{\mu\lambda}+\delta T^{\mu\lambda}$ and the nonminimal coupling term be $\hat{A}^{\lambda}=A^{\lambda}+\delta A^{\lambda}$, where the tensor $\delta T^{\mu\lambda}$ describes the material distribution inside the test particle and $\delta A^{\lambda}$ stands for the test particle's contribution to the coupling term. If we assume that the test particle's trajectory is outside the massive bodies then the background energy-momentum tensor $T^{\mu\lambda}$ vanishes inside as well as near the test particle. Thus, using the identity \eqref{yek} we get
\begin{equation}
\nabla_{\mu}\tau^{\mu\lambda}=\delta f^{\lambda}
\label{do}
\end{equation}
where $\tau^{\mu\lambda}=\sqrt{-g}~\delta T^{\mu\lambda}$ and $\delta f^{\lambda}=\sqrt{-g}~\delta A^{\lambda}$. As we mentioned before, the use of \eqref{do} alone is sufficient to derive the equations of motion of the test particle, and we do not need the filed equations of the theory. In the following sections by using this equation we derive the generalized version of the geodesic equation. Furthermore, we find the generalized Mathisson-Papapetrou equations for describing a spinning test particle.
\section{The single-pole particle}
In this section, we derive the equations of motion for a single-pole test body. Our analysis is based on the integration of the conservation law \eqref{do} over the world tube of the test body. This procedure is independent of a specific choice of energy-momentum tensor for the test particle. In fact, in the four-dimensional diagram the interior of the particle can be considered as a tube-like region (world tube) contained in the 3-dimensional time-like hypersurface $\Sigma(t)$. A representative continuous curve through the tube is parametrized by
 $X^{\lambda}(t)$. Coordinates within the word tube with respect to a coordinate system centered on $X^{\lambda}(t)$ are labeled by $x^{\lambda}$, see
 \cite{papa} for details.

 Equation \eqref{do} can be rewritten as
 \begin{equation}
 \partial_{\mu}\tau^{\mu\lambda}(x)+\Gamma^{\lambda}_{\mu\alpha}(x)\tau^{\mu\alpha}(x)=\delta f^{\lambda}(x)
 \label{cons1}
 \end{equation}
 where $\Gamma^{\lambda}_{\mu\alpha}(x)$ are the Christoffel symbols corresponding to the background metric $g_{\mu\nu}$. Since $g_{\mu\nu}$ changes very little inside the test particle, the Christoffel symbols can be developed in Taylor series around the point $X^{\lambda}(t)$
  \begin{equation}
\Gamma^{\lambda}_{\mu\alpha}(x)=\Gamma^{\lambda}_{\mu\alpha}(X)+\delta x^{\sigma}\partial_{\sigma}\Gamma^{\lambda}_{\mu\alpha}(X)+...
\label{se}
 \end{equation}
where $\delta x^{\sigma}=x^{\sigma}-X^{\sigma}$. The single-pole particle has the simplest internal structure and its dipole as well as higher multipole moments are zero
\begin{equation}
\int_{\Sigma} \delta x^{\sigma} \tau ^{\mu\lambda} d^3x=0, ~~~~\int_{\Sigma} \delta x ^{\sigma} \delta x ^{\alpha} \tau ^{\mu\lambda} d^3x=0, ...
\label{sing0}
\end{equation}
Note that the integrals are taken over the 3-dimensional hyperspace $\Sigma(t)$ which denotes the interior of the test particle at time $t$. Substituting \eqref{se} into \eqref{cons1} and taking into account that we are considering a single-pole particle, we get
 \begin{equation}
 \partial_{\mu}\tau^{\mu\lambda}(x)+\Gamma^{\lambda}_{\mu\alpha}(X)\tau^{\mu\alpha}(x)=\delta f^{\lambda}(x)
 \label{cons2}
 \end{equation}
Integrating this equation over the hypersurface $\Sigma$ and taking into account that $\tau^{\mu\nu}$ are zero outside $\Sigma$, we find
\begin{equation}
\frac{d}{dt}\int \tau^{\lambda 0} d^3x+\Gamma^{\lambda}_{\mu\alpha}(X)\int \tau^{\mu\alpha}d^3x=\int \delta f^{\lambda} d^3x
\label{sing1}
\end{equation}
On the other hand, multiplication of equation \eqref{cons2} by $x^{\nu}$ yields
\begin{equation}
\tau^{\nu\lambda}=\partial_{\mu} (x^{\nu}\tau^{\mu\lambda})+x^{\nu}\Gamma^{\lambda}_{\mu\alpha}(X)\tau^{\mu\alpha}-x^{\nu}\delta f ^{\lambda}
\end{equation}
Integrating this equation over $\Sigma$, we derive
\begin{equation}
\begin{split}
\int \tau ^{\nu\lambda}d^3x=\frac{d}{dt}\int x^{\nu} \tau^{\lambda 0}d^3x+&\Gamma^{\lambda}_{\mu\alpha}(X)\int x^{\nu}\tau^{\mu\alpha}d^3x\\&-\int x^{\nu}\delta f^{\lambda}d^3x
\end{split}
\label{sing2}
\end{equation}
Furthermore, using the requirements \eqref{sing0}, we can write
\begin{equation}
\begin{split}
&\int x^{\nu} \tau^{\lambda 0}d^3x=X^{\nu}(t)\int \tau^{\lambda 0}d^3x\\ &
\int x^{\nu}\tau^{\mu\alpha}d^3x=X^{\nu}(t)\int\tau^{\mu\alpha}d^3x\\&
\int x^{\nu}\delta f^{\lambda}d^3x=X^{\nu}(t)\int \delta f^{\lambda}d^3x
\end{split}
\end{equation}
Substituting these equations into \eqref{sing2} and using equation \eqref{sing1}, one gets
\begin{equation}
\int \tau ^{\nu\lambda}d^3x=\frac{dX^{\nu}}{dt}\int \tau^{\lambda 0}d^3x
\label{sing3}
\end{equation}
With the help of the $\lambda=0$ component, we rewrite \eqref{sing3} as follows
\begin{equation}
\int \tau ^{\nu\lambda}d^3x=\frac{dX^{\nu}}{dt}\frac{dX^{\lambda}}{dt}\int \tau^{0 0}d^3x
\label{sing4}
\end{equation}
Combining equations \eqref{sing1} and \eqref{sing4} we find
\begin{equation}
\begin{split}
\frac{d}{dt}\left(\frac{dX^{\lambda}}{dt}\int \tau^{0 0}d^3x\right)+&\Gamma^{\lambda}_{\mu\alpha}(X)\frac{dX^{\mu}}{dt}\frac{dX^{\alpha}}{dt}\int \tau ^{00}d^3x\\&=\int \delta f^{\lambda}d^3x
\end{split}
\label{sing5}
\end{equation}
On the other hand, the line-element of the curve $X^{\lambda}(t)$ is given by $ds^2=g_{\mu\nu}dX^{\mu}dX^{\nu}$. Also, the corresponding four-velocity is $u^{\lambda}
=\frac{dX^{\lambda}}{ds}$. Thus equation \eqref{sing5} becomes
\begin{equation}
\frac{d}{ds} (m u^{\lambda})+m \Gamma^{\lambda}_{\mu\alpha}u^{\mu}u^{\alpha}=u^0 \int \delta f^{\lambda} d^3x
\label{sing6}
\end{equation}
where $m$ is defined as follows
\begin{equation}
m=\frac{1}{u^0} \int \tau^{00} d^3x
\end{equation}
Multiplying equation \eqref{sing6} by $u_{\lambda}$ and taking into account that $u^{\lambda}\nabla_{\mu} u_{\lambda}=0$, we find
\begin{equation}
\frac{dm}{ds}=u^0 u_{\lambda}\int \delta f^{\lambda} d^3x
\label{sing7}
\end{equation}
In general relativity, the right hand side of this equation is zero and so $m$ is a constant of the motion. In this case $m$ is interpreted as the rest-mass of
the particle \cite{papa}. Substituting \eqref{sing7} into \eqref{sing6} we find
\begin{equation}
m\left(\frac{du^{\lambda}}{ds}+\Gamma^{\lambda}_{\mu\alpha}u^{\mu}u^{\alpha}\right)=(\delta^{\lambda}_{\sigma}-u^{\lambda} u_{\sigma})\Delta^{\sigma}
\label{finalsing}
\end{equation}
here we introduced $\Delta^{\sigma}$ as
\begin{equation}
\Delta^{\sigma}=u^0 \int \delta f^{\sigma} d^3x
\end{equation}
Finally, equation \eqref{finalsing} can be brought into a more compact form, namely
\begin{equation}
m\frac{Du^{\lambda}}{ds}=(\delta^{\lambda}_{\sigma}-u^{\lambda} u_{\sigma})\Delta^{\sigma}
\label{pe1}
\end{equation}
where $\frac{D}{ds}=u^{\mu}\nabla_{\mu}$ is the directional covariant derivative. Equation \eqref{pe1} is the generalization of the geodesic equation. As we expect, orbits of single-pole particles are not the geodesics of the background metric. From \eqref{pe1}, one can immediately read off the additional
contribution to the equations of motion due to the nonzero covariant divergence of the matter energy-momentum tensor. One can interpret the left-hand side of \eqref{pe1} as an extra force - "extra" in comparison to the case $\Delta^{\sigma}=0$ - influencing the test particle motion. Such an extra force due to the nonminimal coupling term on the right-hand side of \eqref{pe1} may provide an interpretation for the observed mass discrepancy in spiral galaxies and clusters of galaxies \cite{de Blok:2001mf,Sanders:2002pf}, or in the context of the so-called Pioneer anomaly \cite{Anderson:2001sg}. For some attempts to construct consistent modified gravity theories that might yield to an appropriate extra force for explaining the aforementioned problems, see \cite{mofi,Bertolami:2007gv} and references therein.

\section{The pole-dipole particle}

In this section, we derive the equations of motion of a pole-dipole particle. The pole-dipole particle is a particle for which all integrals $\int \delta x^{\rho_1}\delta x^{\rho_2}...\delta x^{\rho_n}\tau ^{\mu\nu}d^3x$ with more than one factor $\delta x^{\rho}$ vanish. Remembering that the test particle is a pole-dipole particle, we substitute equation \eqref{se} into
\eqref{cons1} and then integrate over the hypersurface $\Sigma$. The result is
\begin{equation}
\begin{split}
\frac{d}{dt}\int \tau^{\lambda 0} d^3x+&\Gamma^{\lambda}_{\mu\alpha}(X)\int \tau^{\mu\alpha}d^3x+\\&\partial_{\sigma}\Gamma^{\lambda}_{\mu\alpha}(X)\int \delta x^{\sigma}\tau^{\mu\alpha}d^3x=\int \delta f^{\lambda} d^3x
\end{split}
\label{di1}
\end{equation}
Also, multiplying equation \eqref{cons1} by $x^{\nu}$ and integrating over $\Sigma$, it follows that
\begin{equation}
\begin{split}
\int \tau^{\nu\lambda}d^3x&=\frac{d}{dt}\int \delta x^{\nu}\tau^{\lambda 0}d^3x+\frac{dX^{\nu}}{dt}\int\tau^{\lambda 0}d^3x+\\&\Gamma^{\lambda}_{\mu\alpha}(X)\int \delta x^{\nu}\tau^{\mu\alpha}d^3x-\int \delta x^{\nu}\delta f^{\lambda}d^3x
\end{split}
\label{di2}
\end{equation}
Here we have used \eqref{di1} to simplify \eqref{di2}. Another useful equation can be obtained by multiplying equation \eqref{cons2} by $x^{\alpha}x^{\beta}$ and integrating over $\Sigma$. The result is
\begin{equation}
\begin{split}
\frac{dX^{\alpha}}{dt}\int \delta x^{\beta}\tau^{\lambda 0}d^3x+&\frac{dX^{\beta}}{dt}\int \delta x^{\alpha}\tau^{\lambda 0}d^3x=\int \delta x^{\alpha}\tau^{\beta \lambda}d^3x\\&+\int \delta x^{\beta}\tau^{\alpha \lambda}d^3x
\end{split}
\label{di3}
\end{equation}
Equations \eqref{di1}-\eqref{di3} contain enough information to describe the motion of pole-dipole particles. In the following, we will bring these equations into more familiar form. To do so, analogously to \cite{papa} we introduce the quantities $M^{\lambda\alpha\beta}$, $S^{\alpha\beta}$ and $M^{\alpha\beta}$ as follows
\begin{equation}
M^{\lambda\alpha\beta}=-u^0\int \delta x^{\lambda}\tau^{\alpha\beta}d^3x
\end{equation}
\begin{equation}
M^{\alpha\beta}=u^0\int \tau^{\alpha\beta}d^3x
\end{equation}
\begin{equation}
S^{\alpha\beta}=-\frac{1}{u^0}(M^{\alpha\beta 0}-M^{\beta\alpha 0})
\end{equation}
where $S^{\alpha\beta}$ is the total angular momentum or macroscopic spin of the test particle. Furthermore, one should note that $M^{0\alpha\beta}=0$ (because $\delta x^{0}$ is zero) and $M^{\alpha 0 0}=-u^0 S^{\alpha 0}$. Also, we define the new quantity $\Delta^{\alpha\beta}$ as follows
\begin{equation}
\Delta^{\alpha\beta}=u^0 \int \delta x^{\alpha}\delta f^{\lambda} d^3x
\end{equation}
With these definitions the equations \eqref{di1}-\eqref{di3} become
\begin{equation}
\frac{d}{ds}\left(\frac{M^{\lambda 0}}{u^0}\right)+\Gamma^{\lambda}_{\mu\alpha}M^{\mu\alpha}-\partial_{\sigma}\Gamma^{\lambda}_{\mu\alpha}
M^{\sigma\mu\alpha}=\Delta^{\lambda}
\label{di5}
\end{equation}
\begin{equation}
M^{\nu\lambda}=\frac{u^{\nu}}{u^0}M^{\lambda 0}-\frac{d}{ds}\left(\frac{M^{\nu\lambda 0}}{u^0}\right)-\Gamma^{\lambda}_{\mu\alpha}M^{\nu\mu\alpha}-\Delta^{\nu\lambda}
\label{di6}
\end{equation}
\begin{equation}
u^0(M^{\alpha\beta\lambda}+M^{\beta\alpha\lambda})=u^{\alpha} M^{\beta\lambda 0}+u^{\beta}M^{\alpha\lambda 0}
\label{di7}
\end{equation}
Cyclic permutation of the indices in equation \eqref{di7} and subtraction of the second from the combination of the first and the third of the permutations yields
\begin{equation}
M^{\alpha\beta\gamma}=-\frac{1}{2}(S^{\alpha\beta}u^{\gamma}+S^{\alpha\gamma}u^{\beta})+\frac{1}{2}\frac{u^{\alpha}}{u^0}(S^{0 \beta}u^{\gamma}+S^{0 \gamma}u^{\beta})
\label{di8}
\end{equation}
By setting $\lambda=0$ in equation \eqref{di6}, we find
\begin{equation}
M^{\nu 0}=\frac{u^{\nu}}{u^0}M^{0 0}-\frac{DS^{\nu 0}}{ds}-\Gamma^{0}_{\mu\alpha}M^{\nu\mu\alpha}-\Delta^{\nu 0}
\label{di9}
\end{equation}
If we contract \eqref{di9} with $u_{\nu}$ and use the same choice for the mass as in \cite{papa}, namely
\begin{equation}
m=\frac{1}{u^0}\left(M^{\nu 0}+\Gamma^{\nu}_{\gamma\sigma}u^{\gamma}S^{\sigma 0}\right)u_{\nu}
\end{equation}
we obtain
\begin{equation}
m=\frac{1}{(u^{0})^2}\left(M^{00}+\Gamma^{0}_{\mu\nu}S^{\mu 0} u^{\nu}\right)+\frac{u_{\rho}}{u^0}\frac{DS^{\rho 0}}{ds}-\frac{u_{\rho}}{u^0}\Delta^{\rho 0}
\label{mm}
\end{equation}
This mass parameter sometimes called the "kinematical" or "monopole" rest mass of the particle \cite{Semerak:1999qc}. Using equations \eqref{di8},\eqref{di9} and \eqref{mm} the equation \eqref{di6} takes the following form
\begin{equation}
\begin{split}
M^{\nu\lambda}=&m u^{\nu}u^{\lambda}+\frac{u^{\nu}}{u^0}\left(\frac{DS^{\lambda 0}}{ds}-u^{\lambda}u_{\rho}\frac{DS^{\rho 0}}{ds}\right)+\Gamma^{\lambda}_{\mu\alpha}S^{\nu\mu}u^{\alpha}\\&+\frac{1}{2}\frac{DS^{\nu\lambda}}{ds}+\frac{d}{ds}\left(\frac{S^{(\nu 0}u^{\lambda)}}{u^0}\right)\\&-\left(\Delta^{\nu\lambda}+\frac{u^{\nu}}{u^0}\Delta^{\lambda 0}+\frac{u^{\nu}u^{\lambda}}{u^0}u_{\rho}\Delta^{\rho 0}\right)
\end{split}
\label{di10}
\end{equation}
where round brackets in $S^{(\nu 0}u^{\lambda)}$ denote symmetrization. Since $M^{\nu\lambda}$ is symmetric, the antisymmetric part of the right-hand side of \eqref{di6} vanishes. Hence one can immediately verify
\begin{equation}
\begin{split}
\frac{u^{\nu}}{u^0}\frac{DS^{\lambda 0}}{ds}-\frac{u^{\lambda}}{u^0}\frac{DS^{\nu 0}}{ds}+&\frac{DS^{\nu 0}}{ds}=\Delta^{\nu\lambda}-\Delta^{\lambda\nu}+\\&\frac{1}{u^0}(u^{\nu}\Delta^{\lambda 0}-u^{\lambda}\Delta^{\nu 0})
\end{split}
\label{di11}
\end{equation}
After some elementary calculations, this equation can be rewritten as
\begin{equation}
\begin{split}
\frac{DS^{\nu\lambda}}{ds}+&u^{\nu}u_{\rho}\frac{DS^{\lambda\rho}}{ds}-u^{\lambda}u_{\rho}\frac{DS^{\nu\rho}}{ds}=(\Delta^{\nu\lambda}-\Delta^{\lambda\nu})+
\\&u^{\nu}u_{\rho}(\Delta^{\lambda\rho}-\Delta^{\rho\lambda})-u^{\lambda}u_{\rho}(\Delta^{\nu\rho}-\Delta^{\rho\nu})
\end{split}
\label{di12}
\end{equation}
This equation should be compared to equation (5.3) in \cite{papa}. With the help of \eqref{di10}-\eqref{di12}, we derive the following equations which are useful in simplifying \eqref{di5}
\begin{equation}
\begin{split}
M^{\nu 0}=u^0[m u^{\nu}+u_{\rho}&\frac{DS^{\nu\rho}}{ds}-\Gamma^{\nu}_{\gamma\sigma}\frac{u
^{\gamma}}{u^0}S^{\sigma 0}-\\&u_{\rho}(\Delta^{\nu\rho}-\Delta^{\rho\nu})]
\end{split}
\label{di14}
\end{equation}
\begin{equation}
\begin{split}
\Gamma^{\alpha}_{\nu\lambda}&M^{\nu\lambda}=\Gamma^{\alpha}_{\nu\lambda} \left(m u^{\nu}u^{\lambda}+u^{\nu}u_{\rho}\frac{DS^{\lambda\rho}}{ds}\right)+\Gamma^{\alpha}_{\nu\lambda}\Gamma^{\lambda}_{\mu\rho}S^{\nu\mu}u^{\rho}
\\&+\Gamma^{\alpha}_{\nu\lambda}\frac{d}{ds}\left(\frac{S^{\nu 0}u^{\lambda}}{u^0}\right)
-\Gamma^{\alpha}_{\nu\lambda}(\Delta^{\nu\lambda}+u^{\nu}u_{\rho}(\Delta^{\lambda\rho}-\Delta^{\rho\lambda}))
\end{split}
\label{di15}
\end{equation}
By using equations \eqref{di14} and \eqref{di15}, we can bring \eqref{di5} into its final covariant form
\begin{equation}
\begin{split}
\frac{D}{ds}&\left(m u^{\nu}+u_{\rho}\frac{DS^{\nu\rho}}{ds}\right)+\frac{1}{2}S^{\mu\alpha}u^{\sigma}R^{\nu}_{~\alpha\sigma\mu}=\\&\frac{D}{ds}\left(u_{\rho}(\Delta^{\nu\rho}-
\Delta^{\rho\nu})\right)+\Delta^{\nu}+u^0\int \nabla_{\alpha}(\delta x^{\alpha}\delta f^{\nu})d^3x
\end{split}
\label{di16}
\end{equation}
where $R^{\nu}_{~\alpha\sigma\mu}$ is the curvature tensor. Also, note that it is straightforward to show
\begin{equation}\label{dfd}
  u^0\int \nabla_{\alpha}(\delta x^{\alpha}\delta f^{\nu})d^3x\simeq  \Gamma^{\nu}_{\alpha\beta}(X)
\Delta^{\alpha\beta}
\end{equation}
Equation \eqref{di16} should be compared to equation (5.7) in \cite{papa}. In general relativity, the right-hand side of \eqref{di16} is zero \cite{papa}.

 It is important to note that the mass parameter $m$ is not necessarily a constant of motion. To show this in more detail, let us define the "generalized momentum" $p^{\nu}$ as follows
\begin{equation}\label{moment}
    p^{\nu}=m u^{\nu}+\frac{DS^{\nu\alpha}}{ds}u_{\alpha}+2u_{\rho}\Delta^{[\rho\nu]}
\end{equation}
where square brackets denote antisymmetrization. Using this equation and taking into account that $m=u_{\nu}p^{\nu}$, it is straightforward to rewrite equations \eqref{di12} and \eqref{di16} as follows
\begin{equation}\label{finalspin2}
\dot{S}^{\nu\lambda}=2\left(p^{[\nu}u^{
\lambda]}+\Delta^{[\nu\lambda]}\right)
\end{equation}
\begin{equation}\label{finalspin1}
\dot{p}^{\nu}=-\frac{1}{2}S^{\mu\alpha}u^{\sigma}R^{\nu}_{~\alpha\sigma\mu}+\left(\Delta^{\nu}+\Gamma^{\nu}_{\alpha\beta}
\Delta^{\alpha\beta}\right)
\end{equation}
where the dot denotes the covariant derivative with respect to the proper time, "$~\dot{}~"=D/ds$. The main result of this section is embodied in the equations \eqref{finalspin2} and \eqref{finalspin1}. These equations should be compared to the well-known Mathisson-Papapetrou equations for pole-dipole test particles in general relativity. It is obvious that if we assume that there is no coupling between matter and the gravitational fields, then $\Delta^{\alpha}$ and $\Delta^{\alpha\beta}$ are zero and equations \eqref{finalspin2} and \eqref{finalspin1} recover the Mathisson-Papapetrou equations.

Using equation \eqref{finalspin1} one can easily show
\begin{equation}\label{jerm}
    \dot{m}=\left(\frac{D}{ds}\left(S^{\nu\rho}u_{\rho}\right)+2 u_{\rho}\Delta^{[\rho\nu]}\right)\dot{u}_{\nu}+u_{\nu}\left(
    \Delta^{\nu}+\Gamma^{\nu}_{\alpha\beta}
\Delta^{\alpha\beta}\right)
\end{equation}
We can also define another mass parameter by $M^2=p_{\alpha}p^{\alpha}$ which in general is different from $m$. The mass parameter $M$ sometimes called the "dynamical", "total" or "effective" rest mass of the test particle \cite{Semerak:1999qc}. If we contract \eqref{finalspin2} with $p_{\nu}\dot{p}_{\lambda}$, we get
\begin{equation}\label{largem}
\dot{M}=-\frac{\dot{p}_{\lambda}}{M m}\frac{D}{ds}\left(S^{\nu\lambda}p_{\nu}\right)+\frac{M}{m}\left(\Delta^{\nu}+\Gamma^{\nu}_{\alpha\beta}
\Delta^{\alpha\beta}\right)u_{\nu
}
\end{equation}
Thus $M$ is not also necessarily a constant of motion.

It is also necessary to mention that just as in general relativity, equations \eqref{finalspin2} and \eqref{finalspin1} are insufficient to describe the motion of the pole-dipole particle. In fact, we have 14 unknown variables ($x^{\alpha}$, $u^{\alpha}$ and $S^{\alpha\beta}$) and only 11 equations i.e. \eqref{finalspin1}, \eqref{finalspin2} and $u^{\alpha}u_{\alpha}=1$. Note that $u^{\alpha}u_{\alpha}=1$ is the mere constraint due to the choice of parametrization of the trajectories. Thus, we need three more equations to make the model self-consistent.

Let us recall the origin of this apparent inconsistency. The above procedure for determining the motion of test particles consists of choosing a representative point (or a representative world line $X^{\lambda}(t)$) in the particle and taking the moments of equation \eqref{yek} about that point. However, we have not specified how to choose this representative point in the test particle. In the other words, there is an arbitrary step in the above method, and it is the choosing of a representative point in the body. Thus, one can expect that if such a point is not uniquely specified, the motion of the representative point will not be fully determined. In general relativity, the proper specification of such a representative world line yields to a supplementary condition to Mathisson-Papapetrou equations which determines the particle's motion \cite{beig}. The two
most widely used supplementary conditions are the Frenkel condition \cite{frenkel} (for massless particles)
\begin{equation}\label{frenkel}
    S^{\alpha\beta}u_{\beta}=0
\end{equation}
and the Tulczyjew condition \cite{tu} (for massive particles)
\begin{equation}\label{tu}
    S^{\alpha\beta}p_{\beta}=0
\end{equation}

In general relativity, with the help of equations \eqref{jerm} and \eqref{largem} one can immediately verify that $m$ is constant when the Frenkel condition \eqref{frenkel} is assumed, and the mass parameter $M$ is constant for the Tulczyjew condition \eqref{tu}. Here, we do not want to derive the corresponding supplementary equations which may yield to constant parameters $m$ and $M$ in the context of modified gravity theories. Because this may need the exact form of the nonminimal coupling term $\delta f^{\nu}$. However, we shall go into this issue for MOG in the next sections.

\section{Test particle motion in MOG}
\label{tpmim}
In this section, we consider the test particle motion in MOG, and we make some comments about this theory. MOG is an alternative theory of gravity, and it is claimed in the literature that this theory can address the dark matter problem in the spiral galaxies and clusters of galaxies \cite{mofi, mofi1, mofi2}. More specifically, MOG is a scalar-tensor-vector gravity theory which postulates, in addition to the metric tensor, three dynamical scalar gravitational fields $G$, $\mu$ and $\omega$ and also a dynamical massive four-vector gravitational field $\phi^{\mu}$, see Appendix \ref{app} for more detail. The vector field $\phi^{\mu}$ is coupled universally to matter. As we mentioned before, such a coupling between matter and the vector field yields to a nonzero covariant divergence of the ordinary matter energy-momentum tensor, and consequently this coupling leads to a modification of the law of gravitation. Here by the law of gravitation, we mean the gravitational force law between two point particles in the weak-field approximation of the given modified gravity theory. We know that general relativity in the Newtonian limit coincides with Newtonian gravity, but this is not the case for other gravity theories. It should be emphasized that, for a modification to the law of gravitation, it is not necessary to construct a theory with nonminimal coupling terms between matter and gravitational fields. In fact, it is well-known that modified gravity theories which yield to a conserved energy-momentum tensor, can also provide a modification to the law of gravitation. For example see metric $f(R)$ theory \cite{odintsov,Capozziello:2006ph}.

In order to investigate the test particle equations of motion we need the exact form of the conservation law \eqref{yek}. We have found the conservation law in Appendix \ref{app}. The result is (equation \eqref{mog23})
 \begin{equation}\label{mog30}
       \nabla_{ \mu}T^{\mu\nu} =B_{\alpha}^{~\nu}J^{\alpha}+\nabla^{\mu}\left(g_{\mu\alpha}\omega\phi^{\nu}\frac{\partial W(\phi)}{\partial \phi_{\alpha}}-2\omega g^{\nu\rho}\frac{\partial W(\phi)}{\partial g^{\mu\rho}}\right)
\end{equation}
where $B_{\mu\nu}=\nabla_{\mu}\phi_{\nu}-\nabla_{\nu}\phi_{\mu}$, $J^{\alpha}$ is a "fifth force" matter current defined by \eqref{current1} and the self-interaction potential $W(\phi)$ is given by \eqref{potential}. As we discussed, the right-hand side of this equation provides the modification of the law of gravitation. It seems that this theory can provide a wide range of modifications to the law of gravitation. In the other words, by choosing different self-interaction potential $W(\phi)$ one can introduce different modifications. In \cite{mofi}, a special modification has been postulated (see equation (32) in \cite{mofi}). This postulation can be considered as a special choice for $W(\phi)$. However, we show that there is no potential $W(\phi)$ corresponding to the extra force postulated in \cite{mofi}. In the other words, we show that the test particle equation of motion postulated in \cite{mofi} is not compatible with the field equations of MOG. Albeit, as we shall discuss, this issue may not change the main results of MOG.

In this theory, the covariant derivatives of the vector field $\phi^{\mu}$ do not appear in $W(\phi)$. Thus, the most general form for $W(\phi)$ can be given as
 \begin{equation}\label{mog24}
    W(\phi)=\sum_{n} c_n (\phi^{\alpha}\phi_{\alpha})^{n}
 \end{equation}
 where $c_n$ are constant coefficients. For such a potential, it is straightforward to show that the second term on the right-hand side of \eqref{mog30} vanishes. Finally, the conservation law of the ordinary matter energy-momentum tensor in MOG reads
 \begin{equation}\label{mog25}
 \nabla_{ \mu}T^{\mu\nu} =B_{\alpha}^{~\nu}J^{\alpha}
 \end{equation}
 Comparing this equation to \eqref{yek}, we get $\delta f^{\nu}=\sqrt{-g} B_{\alpha}^{~\nu}\delta J^{\alpha}$. In the presence of the test particle, the matter current density is given by $\hat{J}^{\nu}=J^{\nu}+\delta J^{\nu}$. On the other hand, since the particle is moving outside of the massive bodies $J^{\nu}$ is zero near and inside of the test particle. Furthermore, since it is assumed that the matter current density is conserved (i.e. $\nabla_{\alpha}J^{\alpha}=0$), one can immediately infer that $\nabla_{\alpha}\delta J^{\alpha}=0$. Consequently it is straightforward to show that $Q=\int \sqrt{-g}\delta J^0 d^3x$ is a time independent quantity. Hence, we may introduce the matter current density $\delta J^{\alpha}$ as follows (see \cite{weinbergbook})
 \begin{equation}\label{current}
    \delta J^{\alpha}=\frac{Q}{\sqrt{-g}}\int \delta^4(x-X) dx^{\alpha}
 \end{equation}
Therefore, it is easy to show that $\Delta^{\sigma}$ is given by
\begin{equation}\label{del}
    \Delta^{\sigma}=q_5 B^{\sigma}_{~\alpha} u^{\alpha}
\end{equation}
where $q_5=-Q$ is the test particle's fifth force charge. In this theory, the fifth force charge is assumed to be proportional to the particle's mass $m$, such that $q_5=\kappa m$. Finally, using equation \eqref{pe1} the equation of motion of the single-pole particle is given by
\begin{equation}\label{mpe1}
    m\dot{u}^{\lambda}=q_5 B^{\lambda}_{\alpha}u^{\alpha}
\end{equation}
Taking into account equation \eqref{sing7} it is clear that as in general relativity $m$ is a constant of motion, and can be considered as the rest mass of the test particle. Equation of motion \eqref{mpe1} is different from what postulated in \cite{mofi} (equation (31)). In fact as it is obvious from \eqref{mpe1} the scalar field $\omega(x)$, and its derivatives do not appear in the equation of motion. As we mentioned before the vector field $\phi^{\mu}$ is coupled to ordinary matter, thus we expect that the vector field or its derivatives exist in the equation of motion. On the other hand, there is a coupling between $\phi^{\mu}$ and $\omega(x)$, see equation \eqref{mog3}, and consequently one may expect that the scalar field $\omega(x)$ should appear in the equation of motion. However, as it is clear from \eqref{mpe1}, it is not the case in this theory. Furthermore, one can easily show that the test particle action governing the equation of motion \eqref{mpe1} is given by
\begin{equation}\label{mpex1}
S_{TP}=-\int(m+q_5 \phi_{\mu}u^{\mu}) ds
\end{equation}
where $s$ is the proper time along the world line of the test particle. This action is also different from what postulated in equation (30) of \cite{mofi}.

In \cite{mofi}, it has been assumed that, for the vacuum solution around a spherically symmetric point-like mass, $\omega(x)$ is nearly constant. This assumption is also consistent with the numerical solution of the vacuum field equations \cite{moficqg}. With this unnecessary assumption, the postulated equation of motion can be written as
\begin{equation}\label{mpeee1}
    m\dot{u}^{\lambda}=\lambda\omega B^{\lambda}_{\alpha}u^{\alpha}
\end{equation}
where $\lambda$ denotes a coupling constant and its magnitude can be fixed using the experimental observations. Equation \eqref{mpeee1} may still seem different from \eqref{mpe1} because of the presence of constant $\omega$. However, since $\lambda$ has been assumed as an arbitrary constant, one can choose it as $\lambda=-\frac{Q}{\omega}$. Consequently, in this case \eqref{mpeee1} coincides with \eqref{mpe1}, and the main result of this theory related to the dark matter problem will not change.

Now let us consider the spinning particles equations of motion in MOG. It is straightforward to verify that $\Delta^{\alpha\beta}$ is zero in MOG. Therefore, the equations of motion \eqref{finalspin2} and \eqref{finalspin1} can be simplified as follows
\begin{equation}\label{di120}
\dot{S}^{\nu\lambda}=2p^{[\nu}u^{
\lambda]}
\end{equation}
\begin{equation}\label{di160}
\dot{p}^{\nu}=-\frac{1}{2}S^{\mu\alpha}u^{\sigma}R^{\nu}_{~\alpha\sigma\mu}+q_5 B^{\nu}_{~\alpha}u^{\alpha}
\end{equation}
where $p^{\nu}=m u^{\nu}+\dot{S}^{\nu\alpha}u_{\alpha}$. Equation \eqref{di120} is the same as in general relativity, but there exists the extra term $q_5 B^{\nu}_{~\alpha}u^{\alpha}$ on the right-hand side of \eqref{di160} relative to Mathisson-Papapetrou equations. It is obvious from equations \eqref{di120} and \eqref{di160}, which describe the motion of electrically neutral pole-dipole particles in MOG,  that there is a similarity between these equations and the Dixon-Souriau equations which determine the motion of charged pole-dipole particles in the presence of electromagnetic fields in the context of general relativity. Mathematically, the origin of this similarity lies in equation \eqref{mog25}. In fact, using Einstein-Maxwell equations one can easily show that the covariant divergence of the ordinary matter energy-momentum tensor is given by
 \begin{equation}\label{EM}
\nabla_{ \mu}T^{\mu\nu} =F_{\alpha}^{~\nu}J^{\alpha}
\end{equation}
where $F_{\mu\nu}$ is the field strength tensor, $J^{\mu}=(\rho,J^{x},J^{y},J^{z})$ is the current four-vector and $\rho$ is the electric charge density. The close similarity between equations \eqref{mog25} and \eqref{EM} yields to a similarity between equations of motion of test particles in
MOG and Dixon-Souriau equations. However, one should note that equations \eqref{di120} and \eqref{di160} are not exactly similar to the Dixon-Souriau equations. In fact, mathematically, they are similar to the equations of motion of a charged spinning particle without magnetic moment, see equation (9) in \cite{ruffini} .

\section{Conserved quantities}
It is well known that the symmetries of the background space-time may yield to the existence of conserved quantities. For example, in general relativity, for a charged test particle moving under the influence of electromagnetic field, if the electromagnetic field satisfies some consistency conditions, then the space-time symmetries guarantee the existence of some conserved quantities. In this section, we consider the conserved quantities related to the existence of the Killing vectors and the Killing tensors. We recall that a Killing vector $\xi^{\mu}$ and a totally symmetric rank-$n$ Killing tensor $K_{\beta_{1}...\beta_{n}}$ satisfy
\begin{equation}\label{killing}
\begin{split}
&\nabla_{\alpha}\nabla_{\beta}\xi_{\mu}=R_{\beta\mu\alpha}^{~~~~\lambda}\xi_{\lambda}\\&
    \nabla_{(\alpha}\xi_{\beta)}=0
    \end{split}
\end{equation}
and
\begin{equation}\label{killingtensor}
    \nabla_{(\alpha}K_{\beta_{1}...\beta_{n})}=0
\end{equation}
Let us first consider the conserved quantities for the motion of a single-pole particle. The equation of motion is given by \eqref{mpe1}. If we assume that the background geometry possesses a Killing vector $\xi_{\mu}$, then one can straightforwardly show
\begin{equation}\label{conser1}
    \frac{D}{ds}\left(\xi_{\alpha}p^{\alpha}+q_5\xi_{\alpha}\phi^{\alpha}\right)=q_5 u^{\mu}L_{\xi}\phi_{\mu}
\end{equation}
where $L_{\xi}\phi_{\mu}$ is the Lie derivative of the vector field $\phi_{\mu}$ with respect to $\xi$. Thus the quantity
\begin{equation}\label{conser2}
Q_{\xi}=\xi_{\alpha}p^{\alpha}+q_5\xi_{\alpha}\phi^{\alpha}
\end{equation}
is conserved if the vector field $\phi_{\mu}$ is Lie-conserved along $\xi$, i.e. $L_{\xi}\phi_{\mu}=0$. Similarly, one can verify that the component of the velocity along $\xi$
\begin{equation}
    u_{\xi}=\xi_{\alpha}u^{\alpha}
\end{equation}
is conserved if
\begin{equation}
    \xi_{\alpha}B^{\alpha}_{~\beta}=0
\end{equation}
This condition can be generalized to quantities nonlinear in velocities if the space-time has appropriate Killing tensors. Let $K_{\beta_{1}...\beta_{n}}$ be a rank-$n$ Killing tensor satisfying equation \eqref{killingtensor}. Then it is straightforward to show
\begin{equation}\label{qk}
    \dot{Q}_{K}=-\frac{q_{5}}{m}K^{\sigma}_{~(\beta_{1}...\beta_{n-1}}B_{\beta_{n})\sigma}u^{\beta_{1}}...u^{\beta_{n}}
\end{equation}
Thus, $Q_K$ is conserved if
\begin{equation}\label{qkk}
    K^{\sigma}_{~(\beta_{1}...\beta_{n-1}}B_{\beta_{n})\sigma}=0
\end{equation}
Since the metric $g_{\mu\nu}$ is a rank two Killing tensor satisfying trivially the condition \eqref{qkk}, we find conservation of the norm of the test particle's velocity i.e. $g_{\mu\nu}u^{\mu}u^{\nu}$.

Now let us consider the conserved quantities for the motion of a spinning test particle in MOG. Using equations \eqref{jerm}, \eqref{largem} and \eqref{di120}, we find
\begin{equation}\label{jerm2}
\begin{split}
    &\dot{m}=\frac{D}{ds}\left(S^{\alpha\beta}u_{\beta}\right)\dot{u}_{\alpha}\\&
    mM\dot{M}=\frac{D}{ds}\left(S^{\alpha\beta}p_{\beta}\right)\dot{p}_{\alpha}\\&
    S\dot{S}=p^{[\alpha}u^{\beta]}S_{\alpha\beta}
    \end{split}
\end{equation}
where $S$ is the magnitude of the spin defined by  $S^2=\frac{1}{2}S_{\alpha\beta}S^{\alpha\beta}$. Thus, $m$ is a constant of the motion when the Frenkel condition \eqref{frenkel} is assumed and accordingly $M$ is constant when the Tulczyjew condition \eqref{tu} is assumed. Also, the magnitude $S$ of the spin is a constant of the motion for both supplementary conditions. Furthermore, if we assume that the background geometry possesses a killing vector $\xi$, then using equations \eqref{di120} and \eqref{di160} we get
\begin{equation}\label{conser12}
    \frac{D}{ds}\left(\xi_{\alpha}p^{\alpha}+q_5\xi_{\alpha}\phi^{\alpha}+\frac{1}{2}S^{\alpha\beta}\nabla_{\alpha}\xi_{\beta}\right)=q_5 u^{\mu}L_{\xi}\phi_{\mu}
\end{equation}
Thus for a spinning test particle the quantity
\begin{equation}\label{conser120}
Q_{\xi}=\xi_{\alpha}p^{\alpha}+q_5\xi_{\alpha}\phi^{\alpha}+\frac{1}{2}S^{\alpha\beta}\nabla_{\alpha}\xi_{\beta}
\end{equation}
is conserved if the vector field $\phi_{\mu}$ is Lie-conserved along $\xi$.
\section{conclusions}
In this paper, we derived the generalized Mathisson-Papapetrou equations in the realm of modified gravity theories allowing a non-minimal coupling between matter and geometry. In the other words, we derived the equations of motion of the single-pole and the pole-dipole (spinning) test particles using the multipole approximation method. Our results are consistent with the results already found in the literature. In the second part of the paper, using the generalized Mathisson-Papapetrou equations, we analyzed the equations of motion of a pole-dipole particle in the context of MOG. Furthermore, we found some conserved quantities for the motion of a test particle in MOG.

In the context of MOG, our results are different from the results already found in the literature. In fact, the equations of motion for the test particle postulated in MOG \cite{mofi} are not true, and they should be replaced by equation \eqref{mpe1}. However, as we discussed in section \ref{tpmim}, this point does not change the main features of this theory considering the dark matter problem.
\section{acknowledgments}
I would like to thank professor John W. Moffat for valuable comments. This work has been supported by National Elite's Foundation of Iran (Tehran, Iran).
\appendix
\section{Covariant divergence of $T^{\mu\nu}$ in MOG} \label{app}
The action of the theory is given by
―begin{equation}
S=S_{Grav}+S_{\phi}+S_{\chi}+S_{\psi}+S_{\omega}+S_{M}
\label{mog1}
\end{equation}
where $S_{M}$ is the action for ordinary matter and
―begin{equation}
S_{Grav}=\frac{1}{16\pi}\int \sqrt{-g}~ d^4 x ~f(\chi) R
\label{mog2}
\end{equation}
―begin{equation}
S_{\phi}=-\int \sqrt{-g}~ d^4 x ~\omega\left(\frac{1}{4}B^{\mu\nu}B_{\mu\nu}+V_{\phi}(\phi,g_{\mu\nu}, \psi)\right)
\label{mog3}
\end{equation}
―begin{equation}
S_{\chi}=\int \sqrt{-g}~ d^4 x ~\left(\frac{1}{2}g^{\mu\nu}\nabla_{\mu}\chi\nabla_{\nu}\chi-V_{\chi}(\chi)\right)
\label{mog4}
\end{equation}
―begin{equation}
S_{\psi}=\int \sqrt{-g}~ d^4 x ~f(\chi)\left(\frac{1}{2}g^{\mu\nu}\nabla_{\mu}\psi\nabla_{\nu}\psi-V_{\psi}(\psi)\right)
\label{mog5}
\end{equation}
―begin{equation}
S_{\omega}=\int \sqrt{-g}~ d^4 x ~f(\chi)\left(\frac{1}{2}g^{\mu\nu}\nabla_{\mu}\omega\nabla_{\nu}\omega-V_{\omega}(\omega)\right)
\label{mog6}
\end{equation}
where $R$ is the Ricci scalar, $B_{\mu\nu}=\nabla_{\mu}\phi_{\nu}-\nabla_{\nu}\phi_{\mu}$ and the scalar fields $\chi$ and $\psi$ are related to the scalar fields $G$ and $\mu$ of the original paper \cite{mofi} by
\begin{equation}
f(\chi)=\frac{\chi^2}{2}=\frac{1}{G},~~~~~~~~~~~\psi=\ln \mu
\end{equation}
also $V_{\chi}$ and $V_{\psi}$ are related to $V_{G}$ and $V_{\mu}$ by
\begin{equation}
    V_{\chi}=\frac{V_G}{G^3}~~~~~~~~~~~V_{\psi}=\frac{V_{\mu}}{\mu^2}
\end{equation}
In this theory, it has been assumed that $\frac{\partial V_{\phi}}{\partial \nabla_{\gamma}\phi_{\alpha}}=0$. In other words, the covariant derivatives of $\phi^{\alpha}$ do not appear in $V_{\phi}$.

The total energy-momentum tensor is given by
\begin{equation}
   T^{total}_{\mu\nu}=T_{\mu\nu}+T^{\phi}_{\mu\nu}+T^{\chi}_{\mu\nu}+T^{\psi}_{\mu\nu}+T^{\omega}_{\mu\nu}
\end{equation}
where
\begin{equation}\label{newnew}
    T_{\mu\nu}=\frac{-2}{\sqrt{-g}}\frac{\delta S_{M}}{\delta g^{\mu\nu}}~~~~~~~~~~~~~T^{Q}_{\mu\nu}=\frac{-2}{\sqrt{-g}}\frac{\delta S_{Q}}{\delta g^{\mu\nu}}
\end{equation}
and $Q$ can be $\phi$, $\chi$, $\psi$ or $\omega$. Using equations \eqref{mog2}-\eqref{mog6} and \eqref{newnew}, one can easily verify
\begin{equation}\label{mog7}
    T^{\phi}_{\mu\nu}=\omega\left(B_{\mu}^{~\alpha}B_{\nu\alpha}-g_{\mu\nu}\left(\frac{1}{4}B^{\rho\sigma}B_{\rho\sigma}+V_{\phi}\right)+2\frac{\partial V_{\phi}}{\partial g^{\mu\nu}}\right)
\end{equation}
\begin{equation}\label{mog8}
    T^{\chi}_{\mu\nu}=-\left(\nabla_{\mu}\chi\nabla_{\nu}\chi-g_{\mu\nu}\left(\frac{1}{2}\nabla_{\alpha}\chi\nabla^{\alpha}\chi-V_{\chi}\right)\right)
\end{equation}
\begin{equation}\label{mog9}
    T^{\psi}_{\mu\nu}=-f(\chi)\left(\nabla_{\mu}\psi\nabla_{\nu}\psi-g_{\mu\nu}\left(\frac{1}{2}\nabla_{\alpha}\psi\nabla^{\alpha}\psi-V_{\psi}\right)\right)
\end{equation}
\begin{equation}\label{mog10}
    T^{\omega}_{\mu\nu}=-f(\chi)\left(\nabla_{\mu}\omega\nabla_{\nu}\omega-g_{\mu\nu}\left(\frac{1}{2}\nabla_{\alpha}\omega\nabla^{\alpha}\omega-V_{\omega}\right)\right)
\end{equation}
 Furthermore, variation of \eqref{mog1} with respect to $g^{\mu\nu}$, $\phi_{\alpha}$, $\chi$, $\psi$ and $\omega$ yields to the following field equations respectively
 \begin{equation}
 f(\chi)R_{\mu\nu}-\frac{1}{2}f(\chi)R g_{\mu\nu}=(\nabla_{\mu}\nabla_{\nu}-g_{\mu\nu}\square)f(\chi)+8\pi T^{total}_{\mu\nu}
 \label{mog11}
 \end{equation}
 \begin{equation}\label{mog12}
    \omega \nabla_{\mu}B^{\alpha\mu}+\nabla_{\mu}\omega B^{\alpha\mu}+\omega \frac{\partial V_{\phi}}{\partial \phi_{\alpha}}=-J^{\alpha}
 \end{equation}
 \begin{equation}
 \begin{split}
    &\square \chi+V'_{\chi}=\frac{f'(\chi)}{16\pi}+\\&f'(\chi)\left(\frac{1}{2}g^{\mu\nu}(\nabla_{\mu}\psi\nabla_{\nu}\psi+\nabla_{\mu}\omega\nabla_{\nu}\omega)
    -(V_{\psi}+V_{\omega})\right)
 \end{split}
 \label{mog13}
 \end{equation}
 \begin{equation}\label{mog14}
    f(\chi)(\square \psi+V'_{\psi})=-f'(\chi)\nabla_{\gamma}\chi\nabla^{\gamma}\psi-\omega\frac{\partial V_{\phi}}{\partial \psi}
 \end{equation}
 \begin{equation}\label{mog15}
     f(\chi)(\square \omega+V'_{\omega})=-f'(\chi)\nabla_{\gamma}\chi\nabla^{\gamma}\omega-\frac{1}{4}B^{\rho\sigma}B_{\rho\sigma}-V_{\phi}
 \end{equation}
 where prime stands for $H'(x)=\frac{dH}{dx}$ and $J^{\alpha}$ is a "fifth force" matter current defined as
 \begin{equation}\label{current1}
J^{\alpha}=-\frac{1}{\sqrt{-g}}\frac{ \delta S_M}{\delta \phi_{\alpha}}
 \end{equation}
Taking the covariant divergence on both sides of equation \eqref{mog11} yields
 \begin{equation}\label{mog16}
 \begin{split}
    \nabla^{\mu}T_{\mu\nu}=&-\frac{f'(\chi)}{16\pi}R\nabla_{\nu}\chi-(\nabla^{ \mu}T^{\phi}_{\mu\nu}+\nabla^{ \mu}T^{\chi}_{\mu\nu}+\\&\nabla^{ \mu}T^{\psi}_{\mu\nu}+\nabla^{ \mu}T^{\omega}_{\mu\nu})
    \end{split}
 \end{equation}
Note that on purely geometrical grounds, $(\square \nabla_{\mu}-\nabla_{\mu}\square)f=R_{\mu\nu}\nabla^{\nu}f$ and $\nabla^{\mu}G_{\mu\nu}=0$, where $f$ is an arbitrary scalar function and $G_{\mu\nu}$ is the Einstein tensor. On the other hand, taking the covariant divergence of equations \eqref{mog7}-\eqref{mog10} and using the field equations \eqref{mog12}-\eqref{mog15}, we find
 \begin{equation}\label{mog17}
 \begin{split}
   \nabla^{ \mu}T^{\phi}_{\mu\nu}=&B_{\nu\alpha}J^{\alpha}-\nabla_{\nu}\omega\left(\frac{1}{4}B^{\rho\sigma}B_{\rho\sigma}+V_{\phi}\right)+\\&2\nabla^{\mu}\left(\omega\frac{\partial
   V_{\phi}}{\partial g^{\mu\nu}}\right)-\omega\nabla_{\nu}V_{\phi}+\omega B_{\nu\alpha}\frac{\partial V_{\phi}}{\partial \phi_{\alpha}}
   \end{split}
 \end{equation}
 \begin{equation}\label{mog18}
 \begin{split}
  \nabla^{ \mu}&T^{\chi}_{\mu\nu}=-\frac{f'(\chi)}{16\pi}R\nabla_{\nu}\chi- \\& f'(\chi)\nabla_{\nu}\chi\left(\frac{1}{2}g^{\alpha\beta}(\nabla_{\alpha}\psi\nabla_{\beta}\psi+\nabla_{\alpha}\omega\nabla_{\beta}\omega)-(V_{\psi}+V_{\omega})\right)
       \end{split}
 \end{equation}
 \begin{equation}\label{mog19}
     \nabla^{ \mu}T^{\psi}_{\mu\nu}=f'(\chi)\left(\frac{1}{2}\nabla_{\alpha}\psi\nabla_{\alpha}\psi-V_{\psi}\right)\nabla_{\nu}\chi+\omega\frac{\partial V_{\phi}}{\partial \psi}\nabla_{\nu}\psi
 \end{equation}
  \begin{equation}\label{mog20}
  \begin{split}
     \nabla^{ \mu}T^{\omega}_{\mu\nu}=&f'(\chi)\left(\frac{1}{2}\nabla_{\alpha}\omega\nabla_{\alpha}\omega-V_{\omega}\right)\nabla_{\nu}\chi+\\&\left(\frac{1}{4}B^{\rho\sigma}B_{\rho\sigma}
     +V_{\phi}\right)\nabla_{\nu}\omega
     \end{split}
 \end{equation}
 In equation \eqref{mog17} we used the following equation
 \begin{equation}\label{ter}
    B_{\mu}^{~\alpha}\nabla^{\mu}B_{\nu\alpha}-\frac{1}{4}\nabla_{\nu}(B^{\rho\sigma}B_{\rho\sigma})=0
 \end{equation}
Equation \eqref{ter} can be easily verified by using the definition of $B_{\mu\nu}$. Now substituting equations \eqref{mog17}-\eqref{mog20} into equation \eqref{mog16} we obtain
\begin{equation}\label{mog21}
       \nabla^{ \mu}T_{\mu\nu} =B_{\alpha\nu}J^{\alpha}+\left(\omega\nabla_{\alpha}\phi_{\nu}\frac{\partial V_{\phi}}{\partial \phi_{\alpha}}-2\nabla^{\mu}\left(\omega\frac{\partial V_{\phi}}{\partial g^{\mu\nu}}\right)\right)
\end{equation}
On the other hand, taking the covariant divergence of equation \eqref{mog12} and assuming $\nabla_{\alpha}J^{\alpha}=0$, one gets
\begin{equation}\label{mog22}
    \nabla_{\alpha}\left(\omega\frac{\partial V_{\phi}}{\partial \phi_{\alpha}}\right)=0
\end{equation}
With the help of this equation we can rewrite \eqref{mog21} as
\begin{equation}\label{mog23}
       \nabla^{ \mu}T_{\mu\nu} =B_{\alpha\nu}J^{\alpha}+\nabla^{\mu}\left(g_{\mu\alpha}\omega\phi_{\nu}\frac{\partial V_{\phi}}{\partial \phi_{\alpha}}-2\omega\frac{\partial V_{\phi}}{\partial g^{\mu\nu}}\right)
\end{equation}
Finally, we choose the following form for the potential $V_{\phi}$ (see equation (20) in \cite{mofi})
\begin{equation}\label{potential}
V_{\phi}=-\frac{1}{2}e^{2\psi}\phi^{\alpha}\phi_{\alpha}+W(\phi)
\end{equation}
where $W(\phi)$ is the vector field $\phi^{\alpha}$ self-interaction contribution. Substituting this potential into \eqref{potential} we get
\begin{equation}\label{mog23}
       \nabla^{ \mu}T_{\mu\nu} =B_{\alpha\nu}J^{\alpha}+\nabla^{\mu}\left(g_{\mu\alpha}\omega\phi_{\nu}\frac{\partial W(\phi)}{\partial \phi_{\alpha}}-2\omega\frac{\partial W(\phi)}{\partial g^{\mu\nu}}\right)
\end{equation}


\begin{thebibliography}{99}
\bibitem{review}
P. Havas {\it Einstein and the History of General Relativity} ({\it Einstein Studies Vol.} {\bf 1}) ed Howard D and  Stachel J (Boston: Birkh\"{a}user) p 234, (1989).
\bibitem{mat}
M. Mathisson, Acta Phys. Pol. {\bf 6}, 163 (1937).

\bibitem{papa}
A. Papapetrou, Proc.\ Roy.\ Soc.\ London A {\bf 209}, 248 (1951).

\bibitem{Suzuki:1996gm}
S. Suzuki and K. I. Maeda, Phys.\ Rev.\ D {\bf 55}, 4848 (1997).

\bibitem{appof}
T. A. Apostolatos, Class. Quantum Grav. {\bf 13}, 799 (1996).

\bibitem{Semerak:1999qc}
O. Semerak, Mon.\ Not.\ Roy.\ Astron.\ Soc. {\bf 308}, 863 (1999).

\bibitem{Mohseni:2000re}
M. Mohseni and H. R. Sepangi, Class.\ Quant.\ Grav. {\bf 17}, 4615 (2000).

\bibitem{Obukhov:2010kn}
Y. N. Obukhov and D. Puetzfeld, Phys.\ Rev.\ D {\bf 83}, 044024 (2011).

\bibitem{dixon}
W. G. Dixon, Philos. Trans. R. Soc. London A {\bf 277}, 59 (1974).

\bibitem{souriau}
J. M. Souriau, Ann. Inst. Henri Poincareｴ, Sect. A {\bf 20}, 22 (1974).

\bibitem{hojman}
R. Hojman and S. Hojman, Phys. Rev. D {\bf 15}, 2724 (1977).

\bibitem{prasanna}
A. R. Prasanna and K. S. Virbhadra, Phys. Lett. A {\bf 138}, 242 (1989).

\bibitem{ruffini}
D. Bini, G. Gemelli and R. Ruffini, Phys. Rev. D {\bf 61}, 064013 (2000).

\bibitem{Yasskin:1980bu}
P. B. Yasskin and W. R. Stoeger, Phys.\ Rev.\ D {\bf 21}, 2081 (1980).

\bibitem{mofijun}
J. W. Moffat, Phys.\ Rev.\ D {\bf 35}, 3733 (1987).

\bibitem{Puetzfeld:2008xu}
D. Puetzfeld and Y. N. Obukhov, Phys.\ Rev.\ D {\bf 78}, 121501 (2008).

\bibitem{de Blok:2001mf}
W. J. G. de Blok, S. S. McGaugh and V. C. Rubin, Astron.\ J. {\bf 122}, 2396 (2001).

\bibitem{Willbook}
C.~M.~Will, \textit{Theory and Experiment in Gravitational
Physics} (Cambridge University Press, Cambridge, England, 1993).

\bibitem{mofi}
J. W. Moffat, JCAP {\bf 0603}, 004 (2006).

\bibitem{Sanders:2002pf}
R. H. Sanders and S. S. McGaugh, Ann.\ Rev.\ Astron.\ Astrophys. {\bf 40}, 263 (2002).

\bibitem{Anderson:2001sg}
J. D. Anderson, P. A. Laing , E. L. Lau, A. S. Liu, M. M. Nieto and S. G. Turyshev, Phys.\ Rev.\ D {\bf 65}, 082004 (2002).



\bibitem{Bertolami:2007gv}
O. Bertolami, C. G. Bohmer, T. Harko and F. S. N. Lobo, Phys.\ Rev.\ D {\bf 75}, 104016 (2007).

\bibitem{beig}
W. Beiglb\"{o}, Commun. Math. Phys. {\bf 5}, 112 (1964).

\bibitem{frenkel}
J. Frenkel, Z. Phys. {\bf 37}, 243 (1926).

\bibitem{tu}
W. Tulczyjew, Acta Phys. Pol. {\bf 18}, 393 (1959).

\bibitem{mofi1}
J. R. Brownstein and J. W. Moffat, Astrophys.\ J. {\bf 636}, 721 (2006).

\bibitem{mofi2}
J. R. Brownstein and J. W. Moffat, Mon.\ Not.\ Roy.\ Astron.\ Soc. {\bf 367}, 527 (2006).

\bibitem{odintsov}
S. Nojiri and S. D. Odintsov, Phys.\ Rept. {\bf 505}, 59 (2011).

\bibitem{Capozziello:2006ph}
S. Capozziello, V. F. Cardone and A. Troisi, Mon.\ Not.\ Roy.\ Astron.\ Soc. {\bf 375}, 1423 (2007).

\bibitem{weinbergbook}
S. Weinberg, { \it Gravitation and cosmology: principles and applications of the general theory of
relativity} (New York: John Wiley and Sons) p 125, (1972).

\bibitem{moficqg}
J. W. Moffat and V. T. Toth, Class. Quant. Grav. {\bf 26}, 085002 (2009).
\end{thebibliography}
\end{document}